\documentclass[draftcls,onecolumn,11pt]{IEEEtran}
\usepackage{setspace}
\usepackage{cite,epsfig,fleqn}
\usepackage{latexsym}
\usepackage{amsmath,amsfonts,amssymb}
\usepackage{graphicx}
\newcommand{{\cG}}{{\cal G}}
\begin{document}
\title{Performance Analysis of Multiple Antenna Multi-User Detection}
\author{Javad Kazemitabar \quad Hamid Jafarkhani
    \thanks{
       This work was supported in part by an NSF Career Award CCR-0238042 and a Multi-University Research Initiative (MURI), grant \# W911NF-04-1-0224.
       The authors are with the Department of EECS
       at the University of California, Irvine; e-mail:
     \texttt{[skazemit,hamidj]@uci.edu}.}
}

\setcounter{page}{1}
\date{}
\maketitle

\begin{abstract}
We derive the diversity order of some multiple antenna multi-user cancellation and detection schemes. The common property of these detection methods is the usage of Alamouti and quasi-orthogonal space-time block codes. For detecting $J$ users each having $N$ transmit antennas, these schemes require only $J$ antennas at the receiver. Our analysis shows that when having $M$ receive antennas, the array-processing schemes provide the diversity order of $N(M-J+1)$. In addition, our results prove that regardless of the number of users or receive antennas, when using maximum-likelihood decoding we get the full transmit and receive diversities, i.e. $NM$, similar to the no-interference scenario.
\end{abstract}


\IEEEpeerreviewmaketitle

\section{Introduction}

Recently, there has been a lot of attention to multi-user detection schemes with simple receiver structures. Among the simplest ones are those that employ space-time codes \cite{ourjournal,further,nagh}. An orthogonal space-time block code (OSTBC) has linear Maximum-Likelihood (ML) decoding complexity in terms of the number of its symbols \cite{Alamouti,tar99}. This is due to the fact that such a code with $K$ symbols can be modeled as $K$ scalar channels, each bearing information of only one symbol. When two users employing similar OSTBCs, transmit data to the same receiver, it is as if we have $K$ scalar channels each bearing information of two super-imposed symbols. Heuristically, to solve two unknowns (symbols), we need two independent linear combinations of them. In our case this translates to having two antennas at the receiver.
Besides OSTBCs, there are other space-time codes that allow applying the same procedure. We have shown in a recent paper how one can apply multi-user detection (MUD) on any number of users with any number of transmit antennas \cite{ourjournal}. In that work we have used a quasi-orthogonal space-time block code (QOSTBC) and its generalization \cite{quasi}. The benefit of the MUD schemes that employ OSTBC or QOSTBC is that they require very few number of receive antennas. For example, those using Alamouti code \cite{Alamouti} or generalized QOSTBC \cite{ourjournal} require as few as the number of users. Moreover, they provide very simple decoding.

Although there has been a lot of work in this area, there is a lack of performance analysis. To the best of our knowledge, a mathematical calculation of the diversity order of these MUD schemes is missing in the literature. Therefore, we were motivated to find the exact diversity order of these schemes.

In a recent work \cite{newcald} however, the authors provide a mathematical model for calculating the equivalent signal-to-noise-ratio (SNR) of different MUD methods. Their work gives us a tool for analyzing the performance of these schemes. In this paper we will derive the diversity order of all the multiple antenna multi-user detection schemes described in \cite{ourjournal} based on the work in \cite{newcald}. These multi-user schemes include those using Alamouti code for 2, QOSTBC for 4 and generalized QOSTBC for higher number of transmit antennas. In this paper,the diversity order is shown to be equal to $N(M-J+1)$, where $J$ is the number of users and $N$ and $M$ are the number of transmit and receive antennas respectively.

The rest of the paper is structured as follows. Section II reviews the concept of diversity and discusses a few methods of deriving it for a system. In Section III we review the multi-user detection using Alamouti scheme. We then derive the diversity order of that scheme for two users. In Section IV, we review the multi-user detection using QOSTBCs and derive the diversity order for it. Section V concludes the paper.

%
\section{Diversity Order in a Communication Scheme}
Diversity is usually defined as the exponent of the Signal-to-Noise-Ratio (SNR) in the error rate expression, high-SNR scenario,
\begin{equation}
d=-\lim_{{\text{SNR}} \rightarrow \infty}\frac{\text{log } P_e}{\text{log }{\text{SNR}}}
\end{equation}
where $P_e$ represents the probability of decoding error.
One can derive other variants of the diversity definition from the above formula. We mention one that will be used frequently in this work. In \cite{TSE} the authors show that in every open-loop MIMO system, the error event is dominated by {\it Outage}. Outage is the scenario when the instantaneous SNR, due to bad channel realization, is unable to support the desired rate. The result from \cite{TSE} states that
\begin{equation}
\lim_{{\text{SNR}} \rightarrow \infty}\frac{\text{log } P_e}{\text{log }{\text{SNR}}}=\lim_{{\text{SNR}} \rightarrow \infty}\frac{\text{log } P_{out}}{\text{log }{\text{SNR}}}
\end{equation}
Therefore, when finding the diversity order, it is sufficient to know the outage behavior of the system \cite{TSE'sbook}
\begin{equation}
d=\lim_{\epsilon \rightarrow 0^{+}}\frac{\text{log } Pr\left\{{\text{Instantaneous SNR}}<\epsilon\right\}}{\text{log }\epsilon}
\label{diversity}
\end{equation}
\section{Multi-User Detection Using Alamouti}
Consider two users transmitting data simultaneously to a single receiver. Assume also, that they are using the Alamouti scheme. We denote the first user's message by ${\bf c}=$($c_1$, $c_2)^T$, and the second user's message by ${\bf s}$=($s_1$, $s_2)^T$. When using Alamouti the original code transmitted will be in the form of 
$\begin{small}\left(
\begin{array}{cc}
c_1 & c_2\\
-c_2^{*}&c_1^{*}
\end{array}
\right)\end{small}$
and
$\begin{small}\left(
\begin{array}{cc}
s_1 & s_2\\
-s_2^{*}&s_1^{*}
\end{array}
\right)\end{small}$.
 As described in \cite{ourjournal} however, one can derive an equivalent notation as following
\begin{equation}
{\bf r}={\bf H}\cdot {\bf c} + {\bf G}\cdot{\bf s} + {\bf n}
\label{system}
\end{equation}
where ${\bf r}$ has entries ${\bf r}_i$=$[{\bf r}_{1i}~-{\bf r}_{2i}^{*}]^T$ with ${\bf r}_{1i}$ and ${\bf r}_{2i}$ being the signals received at the $i$th receive antenna over two consecutive symbol periods. ${\bf n}$ has a Gaussian distribution with $E[{\bf n}{\bf n}^{*}]=\frac{2}{SNR}{\bf I}$. Also, ${\bf H}$ and ${\bf G}$ are the equivalent channel matrices from the first and second user to the receiver, respectively. Assuming 2 receive antennas, ${\bf H}$ and ${\bf G}$ will have an Alamouti structure as follows
\begin{equation}
\begin{array}{llll}
{\bf H}=\left(\begin{array}{l}{\bf H}_1\\{\bf H}_2\end{array}\right)  & \text{and} & {\bf G}=\left(\begin{array}{l}{\bf G}_1\\{\bf G}_2\end{array}\right)&\\
{\bf H}_i=\left(
\begin{array}{cc}
h_{1i}&h_{2i}\\
-h_{2i}^{*}&h_{1i}^{*}\\
\end{array}
\right)&
\text{and}&
{\bf G}_i=\left(
\begin{array}{cc}
g_{1i}&g_{2i}\\
-g_{2i}^{*}&g_{1i}^{*}\\
\end{array}
\right)&\mbox{ for $i$=1,2} \\
\end{array}
\label{H}
\end{equation}
In order to decode the message of each user, one can use several techniques as mentioned in \cite{newcald,nagh}. The most trivial and computationally complex method is decoding both users together. This method, also known as ML, finds {\bf c} and {\bf s} as follows. 
\begin{equation}
\text{argmax  }p({\bf r}|{\bf c},{\bf s})=\text{argmax  }\frac{1}{\pi^2\sigma^4}exp\left(-\frac{1}{2\sigma^2}\|{\bf r}-{\bf Hc}-{\bf Gs}\|^2\right)
\end{equation}
The second method is Array-Processing (AP) and is sometimes named as Zero-Forcing (ZF) or soft interference cancellation. It requires very little computation and has linear decoding complexity. The following shows the the first step of this decoding method to separate {\bf c} and {\bf s},
\begin{equation}
\begin{array}{l}
\left(
\begin{array}{cc}
{\bf I}_2&-{\bf G}_1{\bf G}_2^{-1}\\
-{\bf H}_2{\bf H}_1^{-1}&{\bf I}_2
\end{array}
\right)
\left(
\begin{array}{l}
{\bf r}_1\\
{\bf r}_2
\end{array}
\right)=\\
\left(
\begin{array}{cc}
{\bf H}'&0\\
0&{\bf G}'
\end{array}
\right)
\left(
\begin{array}{l}
{\bf c}\\
{\bf s}
\end{array}
\right)
+
\left(
\begin{array}{l}
{\bf n}_1'\\
{\bf n}_2'
\end{array}
\right)
\end{array}
\end{equation}
Note that the inverse of the Alamouti matrix is a multiple of its Hermitian and therefore easy to compute.

In what follows, first, we prove a lemma that we use in the calculation of the diversity order.\newline
{\bf{Lemma 1:}}~The following equality is valid for all ${\bf H}$ and ${\bf G}$ matrices of the form (\ref{H}):
\begin{equation}
\begin{small}
\begin{array}{l}
\|{\bf H}\|^2\|{\bf G}\|^2-\|{\bf H^{\dagger}G}\|^2=\\
\left(a_5b_1-a_6b_2-a_7b_3-a_8b_4+a_1b_5+a_2b_6+a_3b_7+a_4b_8-
\frac{2(a_1b_4+a_3b_2+a_4b_1-a_2b_3)(b_1b_8+b_2b_7-b_3b_6+b_4b_5)}{b_1^2+b_2^2+b_3^2+b_4^2}\right)^2+\\
\left(a_6b_1+a_5b_2-a_8b_3+a_7b_4+a_1b_6-a_2b_5+a_3b_8-a_4b_7-
\frac{2(a_1b_4+a_3b_2+a_4b_1-a_2b_3)(-b_1b_7+b_2b_8+b_3b_5+b_4b_6)}{b_1^2+b_2^2+b_3^2+b_4^2}\right)^2+\\
\left(a_7b_1+a_8b_2+a_5b_3-a_6b_4+a_1b_7-a_2b_8-a_3b_5+a_4b_6+
\frac{2(a_1b_4+a_3b_2+a_4b_1-a_2b_3)(-b_1b_6+b_2b_5-b_3b_8-b_4b_7)}{b_1^2+b_2^2+b_3^2+b_4^2}\right)^2+\\
\left(a_8b_1-a_7b_2+a_6b_3+a_5b_4+a_1b_8+a_2b_7-a_3b_6-a_4b_5+
\frac{2(a_1b_4+a_3b_2+a_4b_1-a_2b_3)(b_1b_5+b_2b_6+b_3b_7-b_4b_8)}{b_1^2+b_2^2+b_3^2+b_4^2}\right)^2
\end{array}
\label{square}
\end{small}
\end{equation}
where,
\begin{equation}
\begin{array}{ll}
h_{11}= a_1-ja_2, &h_{21}=-a_3+ja_4, \\
h_{12}=-a_5-ja_6,& h_{22}=-a_7-ja_8\\
g_{11}= b_1+jb_2, &g_{21}= b_3+jb_4,\\
g_{12}= b_5+jb_6,& g_{22}= b_7-jb_8
\end{array}
\end{equation}
{\it\textbf{Proof}}:~Can be checked easily after plugging in the auxilary vairables. 

\subsection{Diversity order of ML method}
Consider the system described in Eq. (\ref{system}) with $M$ receive antennas. When using ML, the receiver finds the codeword that satisfies the minimum distance criterion for the following system
\begin{equation}
\begin{small}
\left( 
\begin{array}{llll}
r_{11}&r_{12}&\cdots &r_{1M}\\
r_{21}&r_{22}&\cdots &r_{2M}\\
\end{array}
\right)=\left(
\begin{array}{llll}
c_1 & c_2 & s_1 & s_2\\
-c_2^{*}&c_1^{*} & -s_2^{*} & s_1^{*}
\end{array}
\right)
\left(
\begin{array}{llll}
h_{11}&h_{12}&\cdots&h_{1M}\\
h_{21}&h_{22}&\cdots&h_{2M}\\
g_{11}&g_{12}&\cdots&g_{1M}\\
g_{21}&g_{22}&\cdots&g_{2M}
\end{array}
\right) + \left( 
\begin{array}{llll}
n_{11}&n_{12}&\cdots&n_{1M}\\
n_{21}&n_{22}&\cdots&n_{2M}
\end{array}
\right)
\end{small}
\end{equation}
The diversity of the above system is equal to the minimum rank of all the difference code matrices times the number of receive antennas \cite{Hamid'sbook}. For the above system this value will be $2M$. For more than two users, the diversity order will remain the same since the minimum rank does not change\footnote{The rank of J concatenated Alamoutis-a 2$J$ $\times$ 2 matrix-is always 2}. A similar argument applies to
any full-rank code designed for N transmit antennas, including
codes designed in \cite{ourjournal}, as our reasoning is independent of $N$. Therefore, in general, the diversity of the ML decoding method is equal to $MN$.
\subsection{Diversity order of the array-processing method with 2 receive antennas}
When there are two Alamouti-equipped transmitters, the effective SNR for user number one when using array-processing (zero-forcing) has been derived in \cite{newcald} to be
\begin{equation}
\text{SNR}_{\text{AP}}=\frac{\|{\bf H}\|^2}{\sigma^2}(1-\|\Lambda\|^2)
\end{equation}
where $\Lambda$ is defined as
\begin{equation}
\Lambda=\frac{{\bf{H}}^{\dagger}\bf{G}}{\|{\bf H}\|\|{\bf G}\|}
\end{equation}
We now apply the formula in Eq. (\ref{diversity}) to derive the diversity order.
\begin{equation}
\begin{array}{l}
d_{AP}=\lim_{\epsilon \rightarrow 0^{+}}\frac{\text{log } Pr\left\{\text{SNR}_{AP}<\epsilon\right\}}{\text{log }\epsilon}=
\lim_{\epsilon \rightarrow 0^{+}}\frac{\text{log } Pr\left\{
\frac{\|{\bf H}\|^2.\|{\bf G}\|^2-\|{\bf H}^{\dagger}{\bf G}\|^2}{\sigma^2\|{\bf G}\|^2}
<\epsilon\right\}}{\text{log }\epsilon}
\end{array}
\end{equation}
We can use (\ref{square}) to simplify the numerator
as shown in Eq. (\ref{inter1}) on top of the next page, where $\underline{\bf b}$=$[ b_1 b_2 \cdots b_8]$. In that equation, conditioned on $\underline{\bf b}$, each of the terms inside the 4 main parentheses is a zero-mean real Gaussian random variable due to independence of $a_i$s. Once divided by the square root of the denominator their variance will become equal to one. 
Moreover, it can be easily checked that these Gaussian random variables are independent. Therefore, the sum of their squares is Chi-square distributed with 4 degrees of freedom and has the following density function
\begin{equation}
f(x)=xe^{-x}  ~~x>0
\end{equation}
For small enough $\epsilon$, 
\begin{equation}
\int_{0}^{\sigma^2\epsilon}f(x)dx=\sigma^4\epsilon^2+O(\sigma^4\epsilon^2)
\label{integral2}
\end{equation}
where $f(x)=O(g(x))$ means there is a positive constant $c$ such that $f(x)\leq c g(x)$ for the desired range of $x$. Since the quantity in Eq. (\ref{integral2}) is independent of $\underline{\bf b}$, its expected value with respect to $\underline{\bf b}$ will remain the same. Therefore, we have
\begin{equation}
d=\lim_{\epsilon \rightarrow 0^{+}}\frac{\text{log}(\sigma^4)+\text{log}(\epsilon^2)}{\text{log}(\epsilon)}=2
\end{equation}
\begin{equation}
\begin{array}{l}
 Pr\left\{
\begin{small}
\begin{array}{c}
\left(a_5b_1-a_6b_2-a_7b_3-a_8b_4+a_1b_5+a_2b_6+a_3b_7+a_4b_8-
\frac{2(a_1b_4+a_3b_2+a_4b_1-a_2b_3)(b_1b_8+b_2b_7-b_3b_6+b_4b_5)}{b_1^2+b_3^2+b_4^2+b_2^2}\right)^2+\\
\left(a_6b_1+a_5b_2-a_8b_3+a_7b_4+a_1b_6-a_2b_5+a_3b_8-a_4b_7-
\frac{2(a_1b_4+a_3b_2+a_4b_1-a_2b_3)(-b_1b_7+b_2b_8+b_3b_5+b_4b_6)}{b_1^2+b_2^2+b_3^2+b_4^2}\right)^2+\\
\left(a_7b_1+a_8b_2+a_5b_3-a_6b_4+a_1b_7-a_2b_8-a_3b_5+a_4b_6+
\frac{2(a_1b_4+a_3b_2+a_4b_1-a_2b_3)(-b_1b_6+b_2b_5-b_3b_8-b_4b_7)}{b_1^2+b_2^2+b_3^2+b_4^2}\right)^2+\\
\left(a_8b_1-a_7b_2+a_6b_3+a_5b_4+a_1b_8+a_2b_7-a_3b_6-a_4b_5+
\frac{2(a_1b_4+a_3b_2+a_4b_1-a_2b_3)(b_1b_5+b_2b_6+b_3b_7-b_4b_8)}{b_1+b_2^2+b_3^2+b_4^2}\right)^2\\
\hline \\
\sigma^2(b_1^2+b_2^2+b_3^2+b_4^2+b_5^2+b_6^2+b_7^2+b_8^2)\\
\end{array}
\end{small}
<\epsilon\right\}  \\
\left. \right. \\
=E_{\bf \underline{b}}\left[ 
Pr\left\{
\begin{small}
\begin{array}{c}
\left(\begin{array}{l}\\ \end{array}a_5b_1-a_6b_2-a_7b_3-a_8b_4+a_1b_5+a_2b_6+a_3b_7+a_4b_8-\right.\\
\left.\frac{2(a_1b_4+a_3b_2+a_4b_1-a_2b_3)(b_1b_8+b_2b_7-b_3b_6+b_4b_5)}{b_1^2+b_3^2+b_4^2+b_2^2}\right)^2+\\
\left(\begin{array}{l}\\  \end{array}a_6b_1+a_5b_2-a_8b_3+a_7b_4+a_1b_6-a_2b_5+a_3b_8-a_4b_7-\right.\\
\left.\frac{2(a_1b_4+a_3b_2+a_4b_1-a_2b_3)(-b_1b_7+b_2b_8+b_3b_5+b_4b_6)}{b_1^2+b_2^2+b_3^2+b_4^2}\right)^2+\\
\left(\begin{array}{l}\\  \end{array}a_7b_1+a_8b_2+a_5b_3-a_6b_4+a_1b_7-a_2b_8-a_3b_5+a_4b_6+\right.\\
\left.\frac{2(a_1b_4+a_3b_2+a_4b_1-a_2b_3)(-b_1b_6+b_2b_5-b_3b_8-b_4b_7)}{b_1^2+b_2^2+b_3^2+b_4^2}\right)^2+\\
\left(\begin{array}{l}\\  \end{array}a_8b_1-a_7b_2+a_6b_3+a_5b_4+a_1b_8+a_2b_7-a_3b_6-a_4b_5+\right.\\
\left.\frac{2(a_1b_4+a_3b_2+a_4b_1-a_2b_3)(b_1b_5+b_2b_6+b_3b_7-b_4b_8)}{b_1+b_2^2+b_3^2+b_4^2}\right)^2\\
\hline\\
{b_1^2+b_2^2+b_3^2+b_4^2+b_5^2+b_6^2+b_7^2+b_8^2}
\end{array}
\end{small}
<\sigma^2\epsilon \right|\left. {\bf \underline{b}} \begin{array}{l} \\ \\ \\ \\ \\ \\ \\ \\ \\ \end{array}\right\}
\right] 
\end{array}
\label{inter1}
\end{equation}
 
{\subsection{The case with more than 2 receive antennas}} Let us now assume the previous system with the exception that there are 3 receive antennas rather than two. For this system we have
\begin{equation}
\begin{array}{l}
{\bf r}_1={\bf H}_1\cdot{\bf c} + {\bf G}_1\cdot{\bf s} + {\bf n}_1 \\
{\bf r}_2={\bf H}_2\cdot{\bf c} + {\bf G}_2\cdot{\bf s} + {\bf n}_2 \\
{\bf r}_3={\bf H}_3\cdot{\bf c} + {\bf G}_3\cdot{\bf s} + {\bf n}_3 \\
\end{array}
\end{equation} 
After applying the array processing algorithm and cancelling the effect of user corresponding to message {\bf s} we get
\begin{equation}
\begin{array}{l}
{\bf r}_1^{'}=\left( \frac{{\bf G}_2^{\dagger}{\bf H}_2}{\|{\bf G}_2\|^2} -\frac{{\bf G}_1^{\dagger}{\bf H}_1}{\|{\bf G}_1\|^2} \right) {\bf c} + {\bf n}_1^{'} \\
{\bf r}_2^{'}=\left( \frac{{\bf G}_3^{\dagger}{\bf H}_3}{\|{\bf G}_3\|^2} -\frac{{\bf G}_1^{\dagger}{\bf H}_1}{\|{\bf G}_1\|^2} \right) {\bf c} + {\bf n}_2^{'} \\
\end{array}
\label{3Rxsimp}
\end{equation}
Conditioned on ${\bf G}_i$s, the noise terms $\bf{n}_1^{'}$ and $\bf{n}_2^{'}$ are correlated Gaussian random variables. Similar statement applies to the new channel matrices $\left( \frac{{\bf G}_2^{\dagger}{\bf H}_2}{\|{\bf G}_2\|^2} -\frac{{\bf G}_1^{\dagger}{\bf H}_1}{\|{\bf G}_1\|^2} \right)$ and $\left( \frac{{\bf G}_2^{\dagger}{\bf H}_2}{\|{\bf G}_2\|^2} -\frac{{\bf G}_1^{\dagger}{\bf H}_1}{\|{\bf G}_1\|^2} \right)$. In \cite{paul} it is shown that in a Rayleigh fading system with receive correlation, like the one we have here, the diversity order will be $NM$ as long as the correlation matrix of the channel is full-rank. Since, \cite{paul} assumes white noise, the equivalent correlation matrix in our case will be correlation matrix of the channel multiplied by the inverse of that of the noise. Clearly, the inverse of the correlation matrix of the noise accounts for the noise-whitening operation. Therefore, if we show that both of these two correlation matrices are full-rank, we can conclude that the system in Eq. (\ref{3Rxsimp}) provides a diversity order of 4 ($N=2, M=2$). The correlation matrix of noise is equal to
\begin{equation}
\left(
\begin{array}{cc}
\left( \frac{\sigma^2}{\|{\bf G}_2\|^2}+\frac{\sigma^2}{\|{\bf G}_1\|^2} \right){\bf I}_2 & \frac{\sigma^2}{\|{\bf G}_1\|^2}{\bf I}_2 \\
\frac{\sigma^2}{\|{\bf G}_1\|^2}{\bf I}_2 & \left( \frac{\sigma^2}{\|{\bf G}_3\|^2}+\frac{\sigma^2}{\|{\bf G}_1\|^2} \right){\bf I}_2
\end{array}
\right)
\label{noisecor2x2}
\end{equation}
where ${\bf I}_2$ is the $2\times2$ identity matrix. This matrix is clearly full-rank for almost (surely) all ${\bf G}_i$ realizations. It remains now to find the correlation matrix of the equivalent channel. Since both lines in Eq. (\ref{3Rxsimp}) represent an Alamouti scheme, we can convert them back into the regular Alamouti representation as follows
\begin{equation}
\begin{array}{l}
{\bf y}_1=\left( \begin{array}{cc}c_1 & c_2 \\ -c_2^{*} & c_1^{*}\end{array} \right) \cdot \left( \begin{array}{l}A_1+jA_2\\ A_3+jA_4 \end{array} \right) + ~noise\\
{\bf y}_2=\left( \begin{array}{cc}c_1 & c_2\\ -c_2^{*} & c_1^{*}\end{array} \right) \cdot \left( \begin{array}{l}B_1+jB_2\\ B_3+jB_4 \end{array} \right) + ~noise
\end{array}
\label{3Rx}
\end{equation} 
where the coefficients are normalized so that the noise terms have unit power. 
Using the SNR result from \cite{newcald} and Eq. (\ref{square}) we can write \\
\begin{equation}
\begin{array}{l}
A_1=\frac{a_5b_1-a_6b_2-a_7b_3-a_8b_4+a_1b_5+a_2b_6+a_3b_7+a_4b_8-
\frac{2(a_1b_4+a_3b_2+a_4b_1-a_2b_3)(b_1b_8+b_2b_7-b_3b_6+b_4b_5)}{b_1^2+b_2^2+b_3^2+b_4^2}}{\sigma\sqrt{b_1^2+\cdots +b_8^2}}\\
A_2=\frac{a_6b_1+a_5b_2-a_8b_3+a_7b_4+a_1b_6-a_2b_5+a_3b_8-a_4b_7-
\frac{2(a_1b_4+a_3b_2+a_4b_1-a_2b_3)(-b_1b_7+b_2b_8+b_3b_5+b_4b_6)}{b_1^2+b_2^2+b_3^2+b_4^2}}{\sigma\sqrt{b_1^2+\cdots +b_8^2}}\\
A_3=\frac{a_7b_1+a_8b_2+a_5b_3-a_6b_4+a_1b_7-a_2b_8-a_3b_5+a_4b_6+
\frac{2(a_1b_4+a_3b_2+a_4b_1-a_2b_3)(-b_1b_6+b_2b_5-b_3b_8-b_4b_7)}{b_1^2+b_2^2+b_3^2+b_4^2}}{\sigma\sqrt{b_1^2+\cdots +b_8^2}}\\
A_4=\frac{a_8b_1-a_7b_2+a_6b_3+a_5b_4+a_1b_8+a_2b_7-a_3b_6-a_4b_5+
\frac{2(a_1b_4+a_3b_2+a_4b_1-a_2b_3)(b_1b_5+b_2b_6+b_3b_7-b_4b_8)}{b_1^2+b_2^2+b_3^2+b_4^2}}{\sigma\sqrt{b_1^2+\cdots +b_8^2}}\\
B_1=\frac{a_9b_1-a_{10}b_2-a_{11}b_3-a_{12}b_4+a_1b_9+a_2b_{10}+a_3b_{11}+a_4b_{12}-
\frac{2(a_1b_4+a_3b_2+a_4b_1-a_2b_3)(b_1b_{12}+b_2b_{11}-b_3b_{10}+b_4b_9)}{b_1^2+b_2^2+b_3^2+b_4^2}}{\sigma\sqrt{b_1^2+\cdots b_4^2+b_9^2+\cdots+b_{12}^2}}\\
B_2=\frac{a_{10}b_1+a_9b_2-a_{12}b_3+a_{11}b_4+a_1b_{10}-a_2b_9+a_3b_{12}-a_4b_{11}-
\frac{2(a_1b_4+a_3b_2+a_4b_1-a_2b_3)(-b_1b_{11}+b_2b_{12}+b_3b_9+b_4b_{10})}{b_1^2+b_2^2+b_3^2+b_4^2}}{\sigma\sqrt{b_1^2+\cdots b_4^2+b_9^2+\cdots+b_{12}^2}}\\
B_3=\frac{a_{11}b_1+a_{12}b_2+a_9b_3-a_{10}b_4+a_1b_{11}-a_2b_{12}-a_3b_9+a_4b_{10}+
\frac{2(a_1b_4+a_3b_2+a_4b_1-a_2b_3)(-b_1b_{10}+b_2b_9-b_3b_{12}-b_4b_{11})}{b_1^2+b_2^2+b_3^2+b_4^2}}{\sigma\sqrt{b_1^2+\cdots b_4^2+b_9^2+\cdots+b_{12}^2}}\\
B_4=\frac{a_{12}b_1-a_{11}b_2+a_{10}b_3+a_9b_4+a_1b_{12}+a_2b_{11}-a_3b_{10}-a_4b_9+
\frac{2(a_1b_4+a_3b_2+a_4b_1-a_2b_3)(b_1b_9+b_2b_{10}+b_3b_{11}-b_4b_{12})}{b_1^2+b_2^2+b_3^2+b_4^2}}{\sigma\sqrt{b_1^2+\cdots b_4^2+b_9^2+\cdots+b_{12}^2}}\\
\end{array}
\end{equation}
The above values are real and imaginary parts of the channel coefficients. Instead of finding the complex correlation matrix we can find the following real correlation matrix 
\begin{equation}
{\bf C}=E\{[\underline{\bf A}|\underline{\bf B}]^T[\underline{\bf A}|\underline{\bf B}]\}
\end{equation}
where $\underline{\bf A}=[A_1~A_2~A_3~A_4]$ and $\underline{\bf B}=[B_1~B_2~B_3~B_4]$.
 It can be easily shown that if ${\bf C}$ is full-rank so will be the complex channel correlation matrix. We already know that $\{A_i\}$ and $\{B_i\}$ are each uncorrelated among themselves. Calculating $E\{A_iB_j\}$ we will have
\begin{equation}
{\bf C}=
\frac{1}{\sigma^2}\left(
\begin{array}{cc}
{\bf I} & {\bf X} \\
{\bf X} & {\bf I}
\end{array}
\right)
\label{chancor}
\end{equation}
where 
\begin{equation}
\begin{array}{l}
X=\frac{1}{\sqrt{(b_1^2+\cdots+b_8^2)(b_1^2+\cdots+b_4^2+b_9^2+\cdots+b_{12}^2)}} 
\begin{small}
\left(
\begin{array}{cccc}
b_5&b_6&b_7&b_8\\
b_6&-b_5&b_8&-b_7\\
b_7&-b_8&-b_5&b_6\\
b_8&b_7&-b_6&-b_5
\end{array}
\right)\cdot
\left(
\begin{array}{cccc}
b_9&b_{10}&b_{11}&b_{12}\\
b_{10}&-b_9&-b_{12}&b_{11}\\
b_{11}&b_{12}&-b_9&-b_{10}\\
b_{12}&-b_{11}&b_{10}&-b_9
\end{array}
\right)
\end{small}
\end{array}
\label{gir}
\end{equation}
From \cite{determinant} we have
\begin{equation}
\begin{array}{l}
det({\bf C})=\frac{1}{\sigma^{16}}det(I-X^T\cdot X)=
\frac{1}{\sigma^{16}}(1-\frac{(b_5^2+\cdots+b_8^2)(b_9^2+\cdots+b_{12}^2)}{(b_1^2+\cdots+b_8^2)(b_1^2+\cdots+b_4^2+b_9^2+\cdots+b_{12}^2)} ){\bf I}\neq 0
\end{array}
\end{equation}
Therefore, the system described in Eq. (\ref{3Rx}) will provide full diversity, i.e. $2\times 2$=4. This means that the described array processing scheme provides a diversity order equal to $N\times(M-J+1)$ for the case of $N=2$, $J=2$, and $M=3$.

We now further inspect the diversity order of the scheme by considering the general case of $M$ receive antennas, while keeping the same number of users and transmit antennas. After canceling the effect of the user corresponding to message ${\bf s}$ we get
\begin{equation}
\begin{array}{ll}
{\bf r}_1^{'}&=\left( \frac{{\bf G}_2^{\dagger}{\bf H}_2}{\|{\bf G}_2\|^2} -\frac{{\bf G}_1^{\dagger}{\bf H}_1}{\|{\bf G}_1\|^2} \right) {\bf c} + {\bf n}_1^{'} \\
{\bf r}_2^{'}&=\left( \frac{{\bf G}_3^{\dagger}{\bf H}_3}{\|{\bf G}_3\|^2} -\frac{{\bf G}_1^{\dagger}{\bf H}_1}{\|{\bf G}_1\|^2} \right) {\bf c} + {\bf n}_2^{'} \\
&\vdots \\
{\bf r}_{M-1}^{'}&=\left( \frac{{\bf G}_M^{\dagger}{\bf H}_{M}}{\|{\bf G}_{M}\|^2} -\frac{{\bf G}_1^{\dagger}{\bf H}_1}{\|{\bf G}_1\|^2} \right) {\bf c} + {\bf n}_{M-1}^{'} \\
\end{array}
\label{MRxsimp}
\end{equation}
We will again form the correlation matrix for noise and the equivalent Alamouti channel coefficients and examine whether they are full-rank. The noise correlation matrix will be 
\begin{equation}
\left(
\begin{array}{cccc}
\left( \frac{\sigma^2}{\|{\bf G}_2\|^2}+\frac{\sigma^2}{\|{\bf G}_1\|^2} \right) & \frac{\sigma^2}{\|{\bf G}_1\|^2}& \cdots & \frac{\sigma^2}{\|{\bf G}_1\|^2} \\
\frac{\sigma^2}{\|{\bf G}_1\|^2} & \left( \frac{\sigma^2}{\|{\bf G}_3\|^2}+\frac{\sigma^2}{\|{\bf G}_1\|^2} \right)& \cdots & \frac{\sigma^2}{\|{\bf G}_1\|^2}\\
\vdots&\vdots&\ddots & \vdots \\
\frac{\sigma^2}{\|{\bf G}_1\|^2}& \frac{\sigma^2}{\|{\bf G}_1\|^2} & \cdots & \left( \frac{\sigma^2}{\|{\bf G}_2\|^2}+\frac{\sigma^2}{\|{\bf G}_{M}\|^2} \right)
\end{array}
\right)\otimes {\bf I}_2
\label{otimes}
\end{equation}
The matrix on the left hand side of the tensor product is full-rank since it has $M-1$ nonzero eigenvalues as following\footnote{The eigenvectors of this matrix are standard unit vectors ${\bf e_1}, {\bf e_2}, \cdots, {\bf e_{M-1}}$.}
\begin{equation}
\frac{\sigma^2}{\|{\bf G}_2\|^2}+\frac{\sigma^2}{\|{\bf G}_1\|^2}, \frac{\sigma^2}{\|{\bf G}_3\|^2}+\frac{\sigma^2}{\|{\bf G}_1\|^2}, \cdots, \frac{\sigma^2}{\|{\bf G}_{M}\|^2}+\frac{\sigma^2}{\|{\bf G}_1\|^2}
\end{equation} 
We should now examine the channel correlation matrix. In the general case of $M$ receive antennas, we will have
\begin{equation}
{\bf C}=\frac{1}{\sigma_2}
\left(
\begin{array}{cccc}
{\bf I} & {\bf X}_{12} & \cdots & {\bf X}_{1(M-1)} \\
{\bf X}_{21} & {\bf I} & \cdots & {\bf X}_{2(M-1)} \\
\vdots & \vdots& \ddots & \vdots \\
{\bf X}_{(M-1)1} & {\bf X}_{(M-1)2}& \cdots & {\bf X}_{(M-1)(M-1)}
\end{array}
\right)
\label{bigC}
\end{equation} 
where ${\bf X}_{ij}={\bf B}_i{\bf B}_j^T$ with 
\begin{equation}
\begin{array}{l}
{\bf B}_i=\frac{1}{\sqrt{b_1^2+\cdots+b_4^2+b_{4i+1}^2+\cdots+b_{4(i+1)}^2}}  
\left(
\begin{small}
\begin{array}{cccc}
b_{4i+1}&b_{4i+2}&b_{4i+3}&b_{4(i+2)}\\
b_{4i+2}&-b_{4i+1}&b_{4(i+2)}&-b_{4i+3}\\
b_{4i+2}&-b_{4(i+2)}&-b_{4i+1}&b_{4i+2}\\
b_{4(i+1)}&b_{4i+3}&-b_{4i+2}&-b_{4i+1}
\end{array}
\end{small}
\right)
\end{array}
\end{equation}
It can be checked easily that ${\bf B}_i\cdot {\bf B}_i^T$=$\frac{b_{4i+1}^2+\cdots+b_{4(i+1)}^2}{b_1^2+\cdots+b_4^2+b_{4i+1}^2+\cdots+b_{4(i+1)}^2}$${\bf I}$=$\beta_i$${\bf I}$. It proves that ${\bf C}$ is full-rank if we can find a $4(M-1) \times 4(M-1)$ matrix ${\bf U}$ such that the rank of
\begin{equation}
{\bf U}^T{\bf C}{\bf U}
\end{equation}
is equal to $4M$. We will try to construct ${\bf U}$ based on the following structure
\begin{equation}
{\bf U}=\left(
\begin{array}{lll}
{\bf u}_1&|\cdots |&{\bf u}_{M-1}
\end{array}
\right)
\end{equation}
where ${\bf u}_i$s are $4(M-1)\times 4$ matrices. The following two lemmas will lead us to construct the matrix ${\bf U}$.\\
{\bf{Lemma} 2:} Given $a_i=\frac{1}{\lambda^{*}+\beta_i-1}$ where $\lambda^{*}$ is a root of $\sum_{i=1}^{M-1}{\frac{\beta_i}{\lambda+\beta_i-1}}=1$ we have
\begin{equation}
C\cdot \left(
\begin{array}{ccccccc}
a_1{\bf B}_1 & | &
a_2{\bf B}_2 & | &
\cdots & | &
a_{M-1}{\bf B}_{M-1}
\end{array}
\right)^T
=\lambda^{*} \left(
\begin{array}{ccccccc}
a_1{\bf B}_1 & | &
a_2{\bf B}_2 & | &
\cdots &| &
a_{M-1}{\bf B}_{M-1}
\end{array}
\right)^T
\end{equation}
{\it\textbf{Proof}}:~See Appendix. 

{\bf{Lemma} 3:} The following equation has $M-1$ distinct non-zero roots
\begin{equation}
\sum_{i=1}^{M-1}{\frac{\beta_i}{\lambda+\beta_i-1}}=1
\end{equation}

{\it\textbf{Proof}}:~See Appendix. 

We name these distinct non-zero roots $\lambda^{*}_1,\cdots,\lambda^{*}_{M-1}$. Let us now define ${\bf u}_i$ vectors by
\begin{equation}
{\bf u}_i=\left(
\begin{array}{ccccccc}
a_{1i}{\bf B}_1^T& | &
a_{2i}{\bf B}_2^T& | &
\cdots & | &
a_{{(M-1)}i}{\bf B}_{M-1}^T
\end{array}
\right)^T
\end{equation}
where $a_{mi}=\frac{1}{\lambda^{*}_i+b_m-1}$ for $i,m=1,\cdots,M-1$. From Lemma 3 and properties of ${\bf B}_j$s it is clear that
\begin{equation}
\begin{array}{l}
{\bf C}\cdot {\bf u}_i=\lambda^{*}_i{\bf u}_i\\
{\bf u}_i^T{\bf u}_i=\sum_j{\beta_ja_{ji}^2}{\bf I}=S_i{\bf I}
\end{array}
\end{equation} 
Also, since $\lambda^{*}_i$s are distinct we have
\begin{equation}
\begin{array}{l}
{\bf u}_i^T{\bf C}{\bf u}_j={\bf u}_i^T\lambda^{*}_j{\bf u}_j=\lambda^{*}_j{\bf u}_i^T{\bf u}_j {~~~~ \mbox{   and}}\\
{\bf u}_i^T{\bf C}{\bf u}_j={\bf u}_i^T{\bf C}^T{\bf u}_j=({\bf C}{\bf u}_i)^T{\bf u}_j=\lambda^{*}_i{\bf u}_i^T{\bf u}_j\\ \Longrightarrow (\lambda^{*}_i-\lambda^{*}_j){\bf u}_i^T{\bf u}_j=0
\\ \Longrightarrow {\bf u}_i^T{\bf u}_j=0 {\mbox{ given }} i\neq j.
\end{array}
\end{equation}
We are now ready to show why ${\bf C}$ is full-rank as follows
\begin{equation}
\begin{array}{l}
\begin{tiny}
\left( 
\begin{array}{c}
{\bf u}_1^T\\
{\bf u}_2^T\\
\vdots \\
{\bf u}_{M-1}^T
\end{array}
\right) \cdot {\bf C} \cdot \left(
\begin{array}{ccccccc}
{\bf u}_1&|&{\bf u}_2&|&\cdots&|&{\bf u}_{M-1}
\end{array}
\right)=
\left( 
\begin{array}{c}
{\bf u}_1^T\\
{\bf u}_2^T\\
\vdots \\
{\bf u}_{M-1}^T
\end{array}
\right) \cdot \left(
\begin{array}{ccccccc}
{\bf C}{\bf u}_1&|&{\bf C}{\bf u}_2&|&\cdots&|&{\bf C}{\bf u}_{M-1}
\end{array}
\right)
\end{tiny}
\\=
{\mbox{diag}}\begin{tiny}(S_1\lambda^{*}_1, S_1\lambda^{*}_1, S_1\lambda^{*}_1, S_1\lambda^{*}_1, S_2\lambda^{*}_2, S_2\lambda^{*}_2, S_2\lambda^{*}_2, S_2\lambda^{*}_2,\cdots, S_{M-1}\lambda^{*}_{M-1}, S_{M-1}\lambda^{*}_{M-1}, S_{M-1}\lambda^{*}_{M-1}, S_{M-1}\lambda^{*}_{M-1})
\end{tiny}
\end{array}
\end{equation}
which is clearly full-rank and it proves the same property for the matrix ${\bf C}$.
Therefore, the channel correlation matrix is full-rank and the provided diversity for the scheme described in Eq. (\ref{MRxsimp}) is $2(M-1)$.

{\subsection{The case with more than 2 users}} Let us now assume the multi-user system with 3 users and 3 antennas at the receiver as follows
\begin{equation}
\begin{array}{l}
{\bf r}_1={\bf H}_1\cdot{\bf c} + {\bf G}_1\cdot{\bf s} + {\bf K}_1\cdot{\bf x} + {\bf n}_1 \\
{\bf r}_2={\bf H}_2\cdot{\bf c} + {\bf G}_2\cdot{\bf s} + {\bf K}_2\cdot{\bf x} + {\bf n}_2 \\
{\bf r}_3={\bf H}_3\cdot{\bf c} + {\bf G}_3\cdot{\bf s} + {\bf K}_3\cdot{\bf x} + {\bf n}_3 \\
\end{array}
\label{3users}
\end{equation}
Once we apply the cancellation technique on the user corresponding to message ${\bf x}$ we get
\begin{equation}
\begin{array}{l}
{\bf r}_1^{'}={\bf K}_1^{-1}{\bf r}_1-{\bf K}_3^{-1}{\bf r}_3=({\bf K}_1^{-1}{\bf H}_1-{\bf K}_3^{-1}{\bf H}_3){\bf c}+({\bf K}_1^{-1}{\bf G}_1-{\bf K}_3^{-1}{\bf G}_3){\bf s}+ {\bf z}_1\\
{\bf r}_2^{'}={\bf K}_2^{-1}{\bf r}_2-{\bf K}_3^{-1}{\bf r}_3=({\bf K}_2^{-1}{\bf H}_2-{\bf K}_3^{-1}{\bf H}_3){\bf c}+({\bf K}_2^{-1}{\bf G}_2-{\bf K}_3^{-1}{\bf G}_3){\bf s}+ {\bf z}_2
\end{array}
\label{3usersimp}
\end{equation}
We note that ${\bf K}_i^{-1}=\frac{{\bf K}_i^{\dagger}}{\|{\bf K}_i\|^2}$. Conditioned on ${\bf K}_i$s, the above system represents a Rayleigh fading channel with 2 users and 2 receive antennas. Therefore, similar to the system in (\ref{3Rxsimp}) all the diversity claims of a 2 user systems (conditionally) apply.\footnote{The only difference is that the noise and the channel coefficients are correlated. However, this will not affect the diversity results since the correlation matrices are exactly like those in Eqs. (\ref{noisecor2x2}) and (\ref{chancor}) and therefore full-rank.} In other words, the diversity order will be equal to 2. Taking the expectation over all $K_i$s will not change this constant value and the diversity will remain 2. Similarly, when having $M$ receive antennas for multi-user detection of 3 users we get the diversity order of $2(M-3+1)$. Using induction on the number of users then, we can prove the following theorem.

{\bf{Theorem 1:}}~Suppose we have $J$ Alamouti-equipped users transmitting to the same receiver in the same frequency band that are time synchronized. Let us also assume that at the receiver we have $M$ antennas and we use array processing as explained in \cite{ourjournal}. The diversity provided to each user will be equal to $2(M-J+1)$.


\section{Multi-User Detection for More than Two Transmit Antennas}
In this section we first briefly explain the scheme in \cite{ourjournal} and then find its diversity order. Suppose, we have two users each with 4 transmit antennas using a QOSTBC. They are synchronously transmitting data to a receiver with two receive antennas as following
\begin{equation}
\begin{small}
\begin{array}{l}
\left(
\begin{array}{l}
r_{11} \\
r_{21} \\
r_{31} \\
r_{41}
\end{array}
\right)=\left(
\begin{array}{llll}
c_1& c_2&c_3&c_4\\
-c_2^{*}&c_1^{*}&-c_4^{*}&c_3^{*}\\
c_3&c_4&c_1&c_2\\
-c_4^{*}&c_3^{*}&-c_2^{*}&c_1^{*}
\end{array}
\right)\cdot \left(
\begin{array}{l}
h_{11}\\
h_{21}\\
h_{31}\\
h_{41}
\end{array}\right) +
\left(
\begin{array}{llll}
s_1& s_2&s_3&s_4\\
-s_2^{*}&s_1^{*}&-s_4^{*}&s_3^{*}\\
s_3&s_4&s_1&s_2\\
-s_4^{*}&s_3^{*}&-s_2^{*}&s_1^{*}
\end{array}
\right)\cdot \left(
\begin{array}{l}
g_{11}\\
g_{21}\\
g_{31}\\
g_{41}
\end{array}\right) +
 \left(
\begin{array}{l}
n_{11}\\
n_{21}\\
n_{31}\\
n_{41}
\end{array}
\right) \\
\left(
\begin{array}{l}
r_{12} \\
r_{22} \\
r_{32} \\
r_{42}
\end{array}
\right)=\left(
\begin{array}{llll}
c_1& c_2&c_3&c_4\\
-c_2^{*}&c_1^{*}&-c_4^{*}&c_3^{*}\\
c_3&c_4&c_1&c_2\\
-c_4^{*}&c_3^{*}&-c_2^{*}&c_1^{*}
\end{array}
\right)\cdot \left(
\begin{array}{l}
h_{12}\\
h_{22}\\
h_{32}\\
h_{42}
\end{array}\right) +
\left(
\begin{array}{llll}
s_1& s_2&s_3&s_4\\
-s_2^{*}&s_1^{*}&-s_4^{*}&s_3^{*}\\
s_3&s_4&s_1&s_2\\
-s_4^{*}&s_3^{*}&-s_2^{*}&s_1^{*}
\end{array}
\right)\cdot \left(
\begin{array}{l}
g_{12}\\
g_{22}\\
g_{32}\\
g_{42}
\end{array}\right) + \left(
\begin{array}{l}
n_{12}\\
n_{22}\\
n_{32}\\
n_{42}
\end{array}
\right)
\end{array} 
\label{QOSTBC}
\end{small}
\end{equation}  
We then define 
\begin{equation}
\begin{array}{lll}
{\bf r}_1=\left(
\begin{array}{c}
r_{11}+r_{31}\\
-r_{21}^{*}-r_{41}^{*}
\end{array}\right)& {\mbox {and}} &
{\bf r}_1^{'}=\left(
\begin{array}{c}
r_{11}-r_{31}\\
-r_{21}^{*}+r_{41}^{*}
\end{array}\right)
\end{array}
\end{equation}
Assuming similar definitions for ${\bf r}_2$ and ${\bf r}_2^{'}$ we will have
\begin{equation}
\begin{array}{lll}
{\bf r}_1={\bf H}_1{\bf c}^{+} + {\bf G}_1{\bf s}^{+} + {\bf n}_1 &  , & {\bf r}_1^{'}={\bf H}_1^{'}{\bf c}^{-} + {\bf G}_1^{'}{\bf s}^{-} + {\bf n}_1^{'}\\
{\bf r}_2={\bf H}_2{\bf c}^{+} + {\bf G}_2{\bf s}^{+} + {\bf n}_2 &  , & {\bf r}_2^{'}={\bf H}_1^{'}{\bf c}^{-} + {\bf G}_1^{'}{\bf s}^{-} + {\bf n}_2^{'} 
\end{array}
\label{QOSTBCsimp}
\end{equation}
where 
\begin{equation}
\begin{array}{lll}
{\bf H}_1=\left(
\begin{array}{cc}
h_{11} + h_{31}& h_{21}+h_{41}\\
-h_{11}^{*}-h_{31}^{*}&h_{21}^{*}+h_{41}^{*}
\end{array}\right) & , & 
{\bf H}_1^{'}=\left(
\begin{array}{cc}
h_{11} - h_{31}& h_{21}-h_{41}\\
-h_{11}^{*}+h_{31}^{*}&h_{21}^{*}-h_{41}^{*}
\end{array}\right)\\
{\bf c}^{+}=\left(
\begin{array}{l}
c_1+c_3\\
c_2+c_4
\end{array}
\right) & , &{\bf c}^{-}=\left(
\begin{array}{l}
c_1-c_3\\
c_2-c_4
\end{array}
\right)
\end{array}
\end{equation}
and the rest of the matrices are defined similarly.

Eq. (\ref{QOSTBCsimp}) reminds us of Eq. (\ref{system}) and (\ref{H}). Using the same array-processing algorithm one can cancel the effect of ${\bf s}$ and get the following
\begin{equation}
\begin{small}
\begin{array}{lll}
{\bf r}^{+}={\bf G}_1^{-1}{\bf r}_1-{\bf G}_2^{-1}{\bf r}_2&=({\bf G}_1^{-1}{\bf H}_1-{\bf G}_2^{-1}{\bf H}_2){\bf c}^{+}+({\bf G}_1^{-1}{\bf n}_1-{\bf G}_2^{-1}{\bf n}_2)&={\bf A}{\bf s}^{+}+{\bf z}\\
{\bf r}^{-}={{\bf G}^{'}}_1^{-1}{\bf r}_1^{'}-{{\bf G}^{'}}_2^{-1}{\bf r}_2^{'}&=({{\bf G}^{'}}_1^{-1}{\bf H}_1-{{\bf G}_2^{'}}^{-1}{\bf H}_2^{'}){\bf c}^{-}+({{\bf G}^{'}}_1^{-1}{\bf n}_1-{{\bf G}_2^{'}}^{-1}{\bf n}_2^{'})&={\bf A}^{'}{\bf c}^{-}+{\bf z}^{'}
\end{array}
\end{small}
\end{equation}
where ${\bf A}$ and ${\bf A}^{'}$ can be shown to be in the form of
\begin{equation}
\begin{array}{ll}
{\bf A}=\left(
\begin{array}{cc}
\alpha_1& \alpha_2\\
-\alpha_2^{*}& \alpha_1^{*}
\end{array} \right)&
{\bf A}^{'}=\left(
\begin{array}{cc}
\alpha_1^{'} & \alpha_2^{'}\\
-{\alpha_2^{'}}^{*}&{\alpha_1^{'}}^{*}
\end{array}\right)
\end{array}
\end{equation}
Conditioned on ${\bf G}_i$ and ${\bf G}_i^{'}$ values, the noise terms will be i.i.d. complex Gaussian random variables. A similar argument applies to $\alpha_i$ and $\alpha_i^{'}$. Now, if we perform the reverse of the conversion in (\ref{QOSTBC})-(\ref{QOSTBCsimp}) we get
\begin{equation}
{\bf R}=\left(
\begin{array}{llll}
c_1& c_2&c_3&c_4\\
-c_2^{*}&c_1^{*}&-c_4^{*}&c_3^{*}\\
c_3&c_4&c_1&c_2\\
-c_4^{*}&c_3^{*}&-c_2^{*}&c_1^{*}
\end{array}
\right)\left(
\begin{array}{c}
\frac{\alpha_1+{\alpha^{'}}_1}{2}\\
\frac{\alpha_2+{\alpha^{'}}_2}{2}\\
\frac{\alpha_1-{\alpha^{'}}_1}{2}\\
\frac{\alpha_2-{\alpha^{'}}_2}{2}
\end{array}
\right)
+ \mbox{i.i.d noise}
\end{equation}
Conditioned on ${\bf G}_i$ and ${\bf G}_i^{'}$ values, the above system is equivalent to a single-user QOSTBC with independent noise and Rayleigh fading channel coefficients. This system clearly provides a diversity order of four\footnote{This is assuming rotated constellation for $c_3$ and $c_4$}, even after taking the expectation. Therefore, in the case of $J=2$ users, $N=4$ transmit, and $M=2$ antennas the diversity order is 4=$N(M-J+1)$.

{\subsection{The case with more than 2 receive antennas}}
Let us consider the above system with the exception that there are three receive antennas instead of two. For this system we have
\begin{equation}
\begin{array}{lll}
{\bf r}_1={\bf H}_1{\bf c}^{+} + {\bf G}_1{\bf s}^{+} + {\bf n}_1 &  , & {\bf r}_1^{'}={\bf H}_1^{'}{\bf c}^{-} + {\bf G}_1^{'}{\bf s}^{-} + {\bf n}_1^{'}\\
{\bf r}_2={\bf H}_2{\bf c}^{+} + {\bf G}_2{\bf s}^{+} + {\bf n}_2 &  , & {\bf r}_2^{'}={\bf H}_2^{'}{\bf c}^{-} + {\bf G}_2^{'}{\bf s}^{-} + {\bf n}_2^{'}\\
{\bf r}_3={\bf H}_3{\bf c}^{+} + {\bf G}_3{\bf s}^{+} + {\bf n}_3 &  , & {\bf r}_3^{'}={\bf H}_3^{'}{\bf c}^{-} + {\bf G}_3^{'}{\bf s}^{-} + {\bf n}_3^{'} 
\end{array}
\label{QOSTBC3Rx}
\end{equation}
After canceling out ${\bf s}$ we get 
\begin{equation}
\begin{small}
\begin{array}{lll}
{\bf r}_1^{+}={\bf G}_2^{-1}{\bf r}_2-{\bf G}_1^{-1}{\bf r}_1&=({\bf G}_2^{-1}{\bf H}_2-{\bf G}_1^{-1}{\bf H}_1){\bf c}^{+}+({\bf G}_2^{-1}{\bf n}_2-{\bf G}_1^{-1}{\bf n}_1)&={\bf A}_1{\bf s}^{+}+{\bf z}_1\\
{\bf r}_1^{-}={{\bf G}^{'}}_1^{-1}{\bf r}_1^{'}-{{\bf G}^{'}}_2^{-1}{\bf r}_2^{'}&=({{\bf G}^{'}}_1^{-1}{\bf H}_1-{{\bf G}_2^{'}}^{-1}{\bf H}_2^{'}){\bf c}^{-}+({{\bf G}^{'}}_1^{-1}{\bf n}_1-{{\bf G}_2^{'}}^{-1}{\bf n}_2^{'})&={\bf A}_1^{'}{\bf c}^{-}+{\bf z}_1^{'}\\
{\bf r}_2^{+}={\bf G}_3^{-1}{\bf r}_3-{\bf G}_1^{-1}{\bf r}_1&=({\bf G}_3^{-1}{\bf H}_3-{\bf G}_1^{-1}{\bf H}_1){\bf c}^{+}+({\bf G}_3^{-1}{\bf n}_3-{\bf G}_1^{-1}{\bf n}_1)&={\bf A}_2{\bf s}^{+}+{\bf z}_2\\
{\bf r}_2^{-}={{\bf G}^{'}}_3^{-1}{\bf r}_3^{'}-{{\bf G}^{'}}_1^{-1}{\bf r}_1^{'}&=({{\bf G}^{'}}_3^{-1}{\bf H}_3-{{\bf G}_1^{'}}^{-1}{\bf H}_1^{'}){\bf c}^{-}+({{\bf G}^{'}}_3^{-1}{\bf n}_3-{{\bf G}_1^{'}}^{-1}{\bf n}_1^{'})&={\bf A}_2^{'}{\bf c}^{-}+{\bf z}_2^{'}
\end{array}
\end{small}
\end{equation}
where
\begin{equation}
\begin{array}{lll}
{\bf A}_1=\left(
\begin{array}{cc}
\alpha_{11}&\alpha_{21}\\
-\alpha_{21}^{*}&\alpha_{11}^{*}
\end{array}
\right)& , & {\bf A}_1^{'}=\left(
\begin{array}{ll}
\alpha_{11}^{'}&\alpha_{21}^{'}\\
-{\alpha_{21}^{'}}^{*}&{\alpha_{11}^{'}}^{*}
\end{array}
\right)\\
{\bf A}_2=\left(
\begin{array}{cc}
\alpha_{12}&\alpha_{22}\\
-\alpha_{22}^{*}&\alpha_{12}^{*}
\end{array}
\right)& , & {\bf A}_2^{'}=\left(
\begin{array}{ll}
\alpha_{12}^{'}&\alpha_{22}^{'}\\
-{\alpha_{22}^{'}}^{*}&{\alpha_{12}^{'}}^{*}
\end{array}
\right)
\end{array}
\end{equation}
Although Gaussian, neither the noise terms, nor the channel fades are uncorrelated. The correlation matrix for the $({\bf z}_1~ {\bf z}_2)^T$ will be equal to 
\begin{equation}
\left(
\begin{array}{cc}
\left( \frac{\sigma^2}{\|{\bf G}_2\|^2}+\frac{\sigma^2}{\|{\bf G}_1\|^2} \right){\bf I}_2 & \frac{\sigma^2}{\|{\bf  G}_1\|^2}{\bf I}_2 \\
\frac{\sigma^2}{\|{\bf G}_1\|^2}{\bf I}_2 & \left( \frac{\sigma^2}{\|{\bf G}_3\|^2}+\frac{\sigma^2}{\|{\bf G}_1\|^2} \right){\bf I}_2
\end{array}
\right)
\label{noisecorQ}
\end{equation}
and for $({\bf z}_1^{'}~{\bf z}_2^{'})^T$ it will be
\begin{equation}
\left(
\begin{array}{cc}
\left( \frac{\sigma^2}{\|{\bf G}_2^{'}\|^2}+\frac{\sigma^2}{\|{\bf G}_1^{'}\|^2} \right){\bf I}_2 & \frac{\sigma^2}{\|{\bf G}_1^{'}\|^2}{\bf I}_2 \\
\frac{\sigma^2}{\|{\bf G}_1^{'}\|^2}{\bf I}_2 & \left( \frac{\sigma^2}{\|{\bf G}_3^{'}\|^2}+\frac{\sigma^2}{\|{\bf G}^{'}_1\|^2} \right){\bf I}_2
\end{array}
\right)
\label{noisecorQ'}
\end{equation}
The correlation matrix of $({\bf A}_1~ {\bf A}_2)$ and $({\bf A}_1^{'}~ {\bf A}_2^{'})$ will be of the form
\begin{equation}
\frac{1}{\sigma^2}\left(
\begin{array}{cc}
{\bf I} & {\bf X} \\
{\bf X} & {\bf I}
\end{array}
\right)
\label{chancor2}
\end{equation}
where ${\bf X}$ is similar to Eq. (\ref{gir})\footnote{The new $b_i$ terms are different, but can be calculated using Eqs. (\ref{3Rx}-\ref{gir}).}.%
%
Clearly, all these correlation matrices are full-rank. Now, similar to the 2-receive antenna case, we can perform the reverse conversion and write the above equation in the following form
\begin{equation}
{\bf R}=\left(
\begin{array}{llll}
c_1& c_2&c_3&c_4\\
-c_2^{*}&c_1^{*}&-c_4^{*}&c_3^{*}\\
c_3&c_4&c_1&c_2\\
-c_4^{*}&c_3^{*}&-c_2^{*}&c_1^{*}
\end{array}
\right)\left(
\begin{array}{cc}
\frac{\alpha_{11}+{\alpha^{'}}_{11}}{2}& \frac{\alpha_{12}+{\alpha^{'}}_{12}}{2}\\
\frac{\alpha_{21}+{\alpha^{'}}_{21}}{2}& \frac{\alpha_{21}+{\alpha^{'}}_{22}}{2}\\
\frac{\alpha_{11}-{\alpha^{'}}_{11}}{2}& \frac{\alpha_{12}-{\alpha^{'}}_{12}}{2}\\
\frac{\alpha_{21}-{\alpha^{'}}_{21}}{2}& \frac{\alpha_{22}-{\alpha^{'}}_{22}}{2}
\end{array}
\right)
+ \mbox{noise}
\label{QOSTBCsimp3}
\end{equation}
The correlation matrix of the new channel and noise terms can be derived via row operations and block-concatenation of the correlation matrices in (\ref{noisecor2x2},\ref{gir}). Therefore, they will also be full-rank and the diversity order of the equivalent scheme shown in (\ref{QOSTBCsimp3}) will be $4\times 2$=8. 

For the general case of $M$ receive antennas, one can perform similar operations and get to the noise and channel correlation matrices like those in (\ref{otimes}) and (\ref{bigC}). After reverse conversion, the equivalent single-user system will look like
\begin{equation}
{\bf R}=\left(
\begin{array}{llll}
c_1& c_2&c_3&c_4\\
-c_2^{*}&c_1^{*}&-c_4^{*}&c_3^{*}\\
c_3&c_4&c_1&c_2\\
-c_4^{*}&c_3^{*}&-c_2^{*}&c_1^{*}
\end{array}
\right)\left(
\begin{array}{cccc}
\frac{\alpha_{11}+{\alpha^{'}}_{11}}{2}& \frac{\alpha_{12}+{\alpha^{'}}_{12}}{2}& \cdots & \frac{\alpha_{1(M-1)}+{\alpha^{'}}_{1(M-1)}}{2}\\
\frac{\alpha_{21}+{\alpha^{'}}_{21}}{2}& \frac{\alpha_{22}+{\alpha^{'}}_{22}}{2}& \cdots &  \frac{\alpha_{2(M-1)}+{\alpha^{'}}_{2(M-1)}}{2}\\
\frac{\alpha_{11}-{\alpha^{'}}_{11}}{2}& \frac{\alpha_{12}-{\alpha^{'}}_{12}}{2}& \cdots & \frac{\alpha_{1(M-1)}-{\alpha^{'}}_{1(M-1)}}{2}\\
\frac{\alpha_{21}-{\alpha^{'}}_{21}}{2}& \frac{\alpha_{22}-{\alpha^{'}}_{22}}{2}& \cdots & \frac{\alpha_{2(M-1)}-{\alpha^{'}}_{2(M-1)}}{2}
\end{array}
\right)
+ \mbox{noise}
\label{QOSTBCsimpM}
\end{equation}
which provides the diversity order of $4(M-1)$ due to the full-rank correlation argument. Therefore, for the case of $J=2$ users, $N=4$ transmit antennas and general $M$ receive antennas, the diversity order will be $N(M-J+1)$.

{\subsection{The case with more than 2 users}} Let us now assume the multi-user system with 3 users and 3 antennas. Let us represent their channel coefficients by $\{h_i\}_{i=1}^4$, $\{g_i\}_{i=1}^4$, and $\{k_i\}_{i=1}^4$ respectively. Naturally the code each of them will transmit will be in the form a QOSTBC.
Now, if we perform the conversion to the Alamouti form, we derive the following equations
\begin{equation}
\begin{array}{lll}
{\bf r}_1={\bf H}_1{\bf c}^{+} + {\bf G}_1{\bf s}^{+} + {\bf n}_1 &  , & {\bf r}_1^{'}={\bf H}_1^{'}{\bf c}^{-} + {\bf G}_1^{'}{\bf s}^{-}+ {\bf K}_1^{'}{\bf x}^{-} + {\bf n}_1^{'}\\
{\bf r}_2={\bf H}_2{\bf c}^{+} + {\bf G}_2{\bf s}^{+} + {\bf n}_2 &  , & {\bf r}_2^{'}={\bf H}_2^{'}{\bf c}^{-} + {\bf G}_2^{'}{\bf s}^{-}+ {\bf K}_2^{'}{\bf x}^{-} + {\bf n}_2^{'} \\
{\bf r}_3={\bf H}_3{\bf c}^{+} + {\bf G}_3{\bf s}^{+} + {\bf n}_3 &  , & {\bf r}_3^{'}={\bf H}_3^{'}{\bf c}^{-} + {\bf G}_3^{'}{\bf s}^{-}+ {\bf K}_3^{'}{\bf x}^{-} + {\bf n}_3^{'} 
\end{array}
\label{QOSTBCsimp3q}
\end{equation}
Once we apply the cancellation technique on the user corresponding to message ${\bf x}$ we get
\begin{equation}
\begin{array}{ll}
{\bf r}_1^{+}={\bf K}_2^{-1}{\bf r}_2-{\bf K}_1^{-1}{\bf r}_1&={\bf A}_1{\bf s}^{+}+{\bf B}_1{\bf x}^{+}+{\bf z}_1\\
{\bf r}_1^{-}={{\bf K}^{'}}_1^{-1}{\bf r}_1^{'}-{{\bf K}^{'}}_2^{-1}{\bf r}_2^{'}&={\bf A}_1^{'}{\bf c}^{-}+{\bf B}_1^{'}{\bf x}^{-}+{\bf z}_1^{'}\\
{\bf r}_2^{+}={\bf K}_3^{-1}{\bf r}_3-{\bf K}_1^{-1}{\bf r}_1&={\bf A}_2{\bf s}^{+}+{\bf B}_2{\bf x}^{+}+{\bf z}_2\\
{\bf r}_2^{-}={{\bf K}^{'}}_3^{-1}{\bf r}_3^{'}-{{\bf K}^{'}}_1^{-1}{\bf r}_1^{'}&={\bf A}_2^{'}{\bf c}^{-}+{\bf B}_2^{'}{\bf x}^{-}+{\bf z}_2^{'}
\end{array}
\end{equation}
Conditioned on ${\bf K}_i$s and ${\bf K}_i^{'}$s, the above system represents a Rayleigh fading channel with 2 users and 2 receive antennas. Therefore, similar to the system in (\ref{3Rxsimp}) all the diversity claims of a 2-user systems (conditionally) apply.\footnote{The only difference is that the noise and the channel coefficients are correlated. However, this will not affect the diversity results since the correlation matrices are exactly like those in Eqs. (\ref{gir}) and (\ref{bigC}) and therefore full-rank.} In other words, the diversity order will be equal to 4. Taking the expectation over all ${\bf K}_i$s and ${\bf K}_i^{'}$s will not change this constant value and the diversity will remain 4. Similarly, when having $M$ receive antennas for multi-user detection of 3 users we get the diversity order of $4(M-3+1)$. Using induction on the number of users then, we can prove the following theorem.

{\bf{Theorem 2:}}~Suppose we have $J$ QOSTBC-equipped users transmitting to the same receiver in the same frequency band that are time synchronized. Let us also assume that at the receiver we have $M$ antennas and we use array processing as explained in \cite{ourjournal}. The diversity provided to each user will be equal to $4(M-J+1)$.

In \cite{ourjournal} we showed how, using $\left(\begin{array}{cc}{\bf A}&{\bf B}\\ {\bf B}&{\bf A}\end{array}\right)$ one can generalize the array processing method to any number of transmit antennas. The trick when $N=2^k$ is to break the system into two systems with $N=2^{k-1}$ and then perform the interference cancellation technique on each of them separately. Then, one can combine them to get the original system. Similar to the method we showed for converting the $N=4$ to $N=2$ systems, one can perform the same diversity analysis on any $N=2^k$-transmit antenna system with $\left(\begin{array}{cc}{\bf A}&{\bf B}\\ {\bf B}&{\bf A}\end{array}\right)$ structure. In addition, the result can be extended to non-power-of-2s with column removal method explained in \cite{ourjournal} to prove the following corollary.

{\bf{Corollary:}}~Assume we have $J$ users each with $N$ transmit antennas using the  $\left(\begin{array}{cc}{\bf A}&{\bf B}\\ {\bf B}&{\bf A}\end{array}\right)$ structure explained above. They are all sending data synchronously to a receiver with $M\geq J$ receive antennas. The diversity of the array processing method explained in \cite{ourjournal} will be equal to $N(M-J+1)$.
\section{Decoding of an interference canceled system}
The algorithm we described in \cite{ourjournal} provides a method to remove unwanted effect of other users and leaves us with a single user system. However, as we noticed in the cases where the number of receive antennas is more than that of users, both the channel and noise coefficients are correlated. We describe the optimal decoding of the system -after interference cancellation- in this section. We prove that this method, which requires ``noise-whitening'' operation, will still keep the separate decoding property of the 2-transmit antenna case. 

As shown, after canceling the first user we have
\begin{equation}
\begin{array}{ll}
{\bf r}_1^{'}=\left( \frac{{\bf G}_2^{\dagger}{\bf H}_2}{\|{\bf G}_2\|^2} -\frac{{\bf G}_1^{\dagger}{\bf H}_1}{\|{\bf G}_1\|^2} \right) {\bf c} + {\bf n}_1^{'} &= {\bf H}_1^{'}{\bf c}+ {\bf n}_1^{'}\\
{\bf r}_2^{'}=\left( \frac{{\bf G}_3^{\dagger}{\bf H}_3}{\|{\bf G}_3\|^2} -\frac{{\bf G}_1^{\dagger}{\bf H}_1}{\|{\bf G}_1\|^2} \right) {\bf c} + {\bf n}_2^{'} &= {\bf H}_2^{'}{\bf c}+ {\bf n}_2^{'}\\
\end{array}
\end{equation}
where the correlation matrix of the noise is
\begin{equation}
{\bf C}_n=\left(
\begin{array}{cc}
\left( \frac{\sigma^2}{\|{\bf G}_2\|^2}+\frac{\sigma^2}{\|{\bf G}_1\|^2} \right){\bf I}_2 & \frac{\sigma^2}{\|{\bf G}_1\|^2}{\bf I}_2 \\
\frac{\sigma^2}{\|{\bf G}_1\|^2}{\bf I}_2 & \left( \frac{\sigma^2}{\|{\bf G}_3\|^2}+\frac{\sigma^2}{\|{\bf G}_1\|^2} \right){\bf I}_2
\end{array}
\right)
\end{equation}
Let us define
\begin{equation}
\begin{array}{ll}
{\bf r}=({\bf r}_1~{\bf r}_2)^T ~~~&~~~
{\bf H}=\left({\bf H}_1^{'} ~ {\bf H}_2^{'} \right)^T
\end{array}
\end{equation}
 Then, the maximum-likelihood decoding metric will be
 \begin{equation}
 \mbox{arg}\min_{\bf c} ({\bf r}-{\bf H}{\bf c})^{\dagger}{\bf C_n}^{-1}({\bf r}-{\bf H}{\bf c}). 
 \end{equation}
 It can be shown \cite{inverse} that 
 \begin{equation}
 {\bf C}_n^{-1}=\left(
 \begin{array}{cc}
 x{\bf I}&y{\bf I}\\
 y{\bf I}&t{\bf I}
 \end{array}
 \right)
 \end{equation}
 where $x,y,t$ are real numbers.
 Therefore, the ML criterion will be to minimize
 \begin{equation}
 \begin{array}{l}
 ({\bf r}_1^{\dagger}-{\bf c}^{\dagger}{\bf H_1}^{\dagger}~{\bf r}_2^{\dagger}-{\bf c}^{\dagger}{\bf H_2}^{\dagger})\left(
 \begin{array}{cc}
 x{\bf I}&y{\bf I}\\
 y{\bf I}&t{\bf I}
 \end{array}
 \right)\left(
 \begin{array}{c}
 {\bf r}_1-{\bf H}_1{\bf c}\\
 {\bf r}_2-{\bf H}_2{\bf c}
 \end{array}\right)\\
 =x\|{\bf r}_1-{\bf H}_1{\bf c}\|^2+ t\|{\bf r}_2-{\bf H}_2{\bf c}\|^2+2yRe\{({\bf r}_1^{\dagger}-{\bf c}^{\dagger}{\bf H}_1^{\dagger})({\bf r}_2-{\bf H}_2{\bf c})\}\\
 =x\|{\bf r}_1-{\bf H}_1{\bf c}\|^2+ t\|{\bf r}_2-{\bf H}_2{\bf c}\|^2+2yRe\{{\bf r}_1^{\dagger}{\bf r}_2-{\bf r}_1^{\dagger}{\bf H}_2{\bf c}-{\bf c}^{\dagger}{\bf H}_1^{\dagger}{\bf r}_2+{\bf c}^{\dagger}{\bf H}_1^{\dagger}{\bf H}_2{\bf c}\}
 \end{array}
 \label{ML}
 \end{equation}
 The only part in the above equation that could generate cross-terms and therefore cause non-separate decoding is ${\bf c}^{\dagger}{\bf H}_1^{\dagger}{\bf H}_2{\bf c}$. Before we expand this term, we note that ${\bf H}_1^{\dagger}{\bf H}_2$ is in the form of an Alamouti matrix and can be written as
 \begin{equation}
 \left(\begin{array}{cc}h_1&h_2\\ -h_2^{*}&h_1^{*}\end{array}\right)
 \end{equation}
  Having that in mind the last term in Eq. (\ref{ML}) can be written as
 \begin{equation}
 \begin{array}{ll}
Re\{h_1|c_1|^2+h_1^{*}|c_2|^2+h_2c_1^{*}c_2-h_2^{*}c_2^{*}c_1\}&=Re\{h_1\}(|c_1|^2+|c_2|^2)+2Re\{j\cdot Im\{h_2c_1^{*}c_2\}\}\\
&=Re\{h_1\}(|c_1|^2+|c_2|^2)
 \end{array}
 \end{equation}
 which clearly does not have any cross-terms and therefore $c_1$ and $c_2$ can be decoded separately.
 \section{conclusion}
We derived the diversity order of some multiple antenna multi-user cancellation and detection schemes. The common property of these detection methods is the usage of Alamouti and quasi-orthogonal space-time block codes. For detecting $J$ users each having $N$ transmit antennas, these schemes require only $J$ antennas at the receiver. Our analysis showed that when having $M$ receive antennas, the array-processing scheme provides the diversity order of $N(M-J+1)$. 
In addition, we proved that regardless of the number of users or receive antennas, when using maximum-likelihood decoding we get the full transmit and receive diversities, i.e. $NM$, similar to the no-interference scenario.
\section*{Appendix}

{\textbf {Proof of Lemma 2}}: Plugging in ${\bf C}$ by 
\begin{equation}
{\bf C}=\left(
\begin{array}{cccc}
{\bf I} & {\bf B}_1{\bf B}_2^T & \cdots & {\bf B}_1{\bf B}_{M-1}^T \\
{\bf B}_2{\bf B}_1^T & {\bf I} & \cdots & {\bf B}_2{\bf B}_{M-1}^T \\
\vdots & \vdots& \ddots & \vdots \\
{\bf B}_{M-1}{\bf B}_1^T & {\bf B}_{M-1}{\bf B}_2^T& \cdots & {\bf I}
\end{array}
\right)
\end{equation}
we get
\begin{equation}
{\bf C} \cdot \left(
\begin{array}{c}
a_1{\bf B}_1\\
a_2{\bf B}_2\\
\vdots \\
a_{M-1}{\bf B}_{M-1}\
\end{array}
\right)=\left(
\begin{array}{c}
(a_1+ a_2\beta_2+\cdots+ a_{M-1}\beta_{M-1}){\bf B}_1 \\
(a_1\beta_1+ a_2+\cdots+ a_{M-1}\beta_{M-1}){\bf B}_2 \\
\vdots \\
(a_1\beta_1+ a_2\beta_2+\cdots+ a_{M-1}){\bf B}_{M-1} 
\end{array}
\right)
\end{equation}
and solving for $a_i$ and $\lambda$ we get
\begin{equation}
\begin{array}{ll}
a_i = \frac{\sum_{i=1}^{M-1}{a_i\beta_i}}{\lambda+\beta_i-1} &
\mbox{for }i=1, 2,\cdots, M-1  
\end{array}
\end{equation}
We can always normalize $a_i$ coefficients such that ${\sum_{i=1}^{M-1}{a_i\beta_i}}=1$. Therefore,
\begin{equation}
\begin{array}{lll}
a_i=\frac{1}{\lambda+\beta_i-1} &
\mbox{and}&
\sum_{i=1}^{M-1}{a_i\beta_i}=\sum_{i=1}^{M-1}{\frac{\beta_i}{\lambda+\beta_i-1}}=1
\end{array}
\end{equation}
which proves the lemma.~~~~~~~~~~~~~~~~~~~~~~~~~~~~~~~~~~~~~~~~~~~~~~~~~~~~~~~~~~~~~~~~~~~~~~~~~~~~~~~~~~~~~~~~~~~~~~~~~~~~~~~~~~~~~~~~~~~~~$\square$

{\textbf {Proof of Lemma 3}}: It is clear why none of the roots can be zero. Because, if it is so, we will have
$\sum_{i=1}^{M}{\frac{\beta_i}{\beta_i-1}}=1$ which is impossible since $\beta_i-1<0$ and $\beta_i>0$ by definition. Also, from the definition we know $\beta_i$s are distinct. Therefore, without loss of generality we can assume $\beta_1 < \cdots <\beta_{M-1}$. It will then be easy to show that $f(\lambda)$=$\sum_{i=1}^{M-1}{\frac{\beta_i}{\lambda+\beta_i-1}}$ is monotonic over the following $M-1$ intervals
\begin{equation}
(1-\beta_{M-1},1-\beta_{M-2}),\cdots,(1-\beta_{2}~,~1-\beta_{1})~,~(1-\beta_1,\infty)
\end{equation}
For the first $M-2$ intervals, $f(\lambda)$ takes all the values from $-\infty$ to $+\infty$. For the last interval, it takes $\infty$ when $\lambda$ is at the proximity of $1-\beta_1$ and 0 when $\lambda$ goes to $\infty$. Therefore, it takes the value of 1 in all of these $M-1$ intervals exactly once, which proves the lemma.
~~~~~~~~~~~~~~~~~~~~~~~~~~~~~~~~~~~~~~~~~~~~~~~~~~~~~~~~~~~~~~~~~~~~~~~~~~~~~~~~~~~~~~~~$\square$
 \section*{Acknowledgement}
 The authors would like to thank Seyed Jalil Kazemitabar from Sharif University of Technology for his useful comments.

\end{document}